# Unusual Scaling Laws of the Band Gap and Optical Absorption of Phosphorene Nanoribbons

Vy Tran[†] and Li Yang[†*]

[†] *Department of Physics, Washington University, St. Louis, Missouri, 63130, USA*

\* E-mail: lyang@physics.wustl.edu

**ABSTRACT:** We report the electronic structure and optical absorption spectra of monolayer black phosphorus (phosphorene) nanoribbons (PNRs) via first-principles simulations. The band gap of PNRs is strongly enhanced by quantum confinement. However, differently oriented PNRs exhibit distinct scaling laws for the band gap vs the ribbon width ($w$). The band gaps of armchair PNRs scale as $1/w^2$, while zigzag PNRs exhibit a $1/w$ behavior. These distinct scaling laws reflect a significant implication of the band dispersion of phosphorene: electrons and holes behave as classical particles along the zigzag direction, but resemble relativistic particles along the armchair direction. This unexpected merging of classical and relativistic properties in a single material may produce novel electrical and magnetotransport properties of few-layer black phosphorus and its ribbon structures. Finally, the respective PNRs host electrons and holes with markedly different effective masses and optical responses, which are suitable for a wide range of applications.

Graphene-inspired two-dimensional (2D) structures have garnered tremendous interest in fundamental science and have inspired broad applications.[1-8] These layered structures can be etched or patterned along a specific lattice direction, forming one-dimensional (1D) strips, called nanoribbons. Graphene nanoribbons (GNRs) and MoS$_2$ nanoribbons are quintessential examples of these 1D strips.[9-11] Because of unique quantum confinement and edge effects, nanoribbons exhibit many exploitable electrical, optical, and magnetic properties.[12-16] Recently, few-layer black phosphorus (phosphorene), a novel 2D direct band gap semiconductor, was successfully exfoliated with promising electric and optical properties.[17-20] Although phosphorene nanoribbons (PNRs) have yet to be fabricated, the research history of graphene and other 2D materials strongly suggests that theoretical predictions of the electronic structures and optical responses of PNRs will be essential for informing and motivating future research of phosphorene-based nanoelectronics.

Unlike other widely studied 2D structures, the band structure, electrical conductivity, and optical responses of few-layer black phosphorus are all highly anisotropic.[18, 21-23] These anisotropies can be associated with the orientation of PNRs, and can be used to dramatically tune the material's transport behaviors and even unexpected electronic properties. This is a distinguishing feature of PNRs that makes them particularly promising for potential applications. Moreover, as seen in other nanoribbons,[24-26] the relaxations and passivations of edge structures will provide additional ways of modifying the PNR's electronic structure.

In this work, we employ first-principles simulations to study the electronic structure and optical absorption spectra of two typical types of PNRs, armchair PNRs (APNRs) and zigzag PNRs (ZPNRs). All of the studied PNRs exhibit a nearly direct band gap, whose size is strongly enhanced by quantum confinement. However, scaling laws of the band gap according to the ribbon width ($w$) are distinct for different ribbon orientations. APNRs exhibit the usual $1/w^2$ relation; the band gaps of ZPNRs, however, scale like $1/w$. This discrepancy arises from markedly-dissimilar behaviors of both the electrons and holes in the parent material, monolayer phosphorene. The zigzag direction hosts free carriers that behave like heavy, classical particles. The armchair direction, on the other hand, hosts relativistic particles with small rest mass. This unexpected merging of classical and relativistic properties in a single material may produce novel transport and magnetic properties of few-layer black phosphorus. Finally, we examine the effect that the lattice orientation has on the effective masses and optical responses. For instance, it is shown that APNRs exhibit an optically-active direct band gap while the optical transitions of ZPNRs are inactive around the band gap due to symmetry-forbidden transitions.

The top views of atomic structures of typical PNRs are presented in Figures 1 (a) and (b), respectively. The edge dangling bonds are fully passivated by hydrogen atoms to stabilize the structures and to quench edge states, which usually reside inside the band gap.[27] Following the conventions used for GNRs,[28] we identify the PNR structures by their number of zigzag chains or P-P dimmers, respectively, along the ribbon orientations. For example, the structure shown in Figure 1 (a) has 7 zigzag chains and is thus indexed as 7-ZPNR; the structure in Figure 1 (b) has 8 P-P dimmers within a unit cell and it is indexed as 8-APNRs.

Our studied structures are fully relaxed according to the force and stress calculated by density functional theory (DFT) with the Perdew, Burke and Ernzerhof (PBE) functional.[29] The plane-wave cutoff is set to be 25 Ry by using norm-conserving pseudopotentials.[30] The k-point sampling is 1x1x16 for electronic-structure calculations and 1x1x240 for single-particle optical absorption spectra without including excitonic effects. Because of the depolarization effect,[31, 32] only the incident light polarized along the ribbon orientation is considered for calculations of optical spectra in this work.

**RESULTS AND DISCUSSION**

The DFT-calculated band structures of two typical PNRs are presented in Figure 2. Their band gaps are significantly enhanced due to quantum confinement. For example, the 6-ZPNR in Figure 2 (a) with a ribbon-width of 1.1 nm exhibits a band gap of 2.0 eV, and the 5-APNR in Figure 2 (b) with a width of 0.8 nm exhibits a band gap of 1.2 eV. Both band gaps are larger than the bulk value (around 0.85 eV). Identifying these band gaps as being direct or indirect proves to be subtle. All of the APNRs display a direct band gap at the Γ point. However, the maximum of valence band of ZPNRs is located slightly away from the Γ point, as is shown in the subfigure in the right-bottom corner of Figure 2 (a). This indicates that ZPNRs are not perfect direct-gap semiconductors, at least at the DFT/PBE level. This is consistent with the DFT/PBE-calculated band structure of monolayer black phosphorus, which exhibits a slight indirect band gap with the same features.[21] This is has not been directly addressed in previous studies because the energy difference between top of valence band and the local maximum at the Γ point is so small (less than 20 meV). Moreover, we find that this energy difference is sensitive to strain conditions and environments, and it may be difficult to observe this energy difference experimentally. Therefore, given this small and delicate energy difference, we regard phosphorene and PNRs as direct-gap semiconductors in this work.

The highly anisotropic band dispersion in monolayer phosphorene makes band structures of ZPNRs wholly distinct from APNRs. As shown in Figure 2 (a), the highest-energy valence band of 6-ZPNRs is extremely flat, with a heavy effective mass (3.2 $m_0$); while that of the 5-APNR is substantially dispersive, with a light effective mass (0.2 $m_0$). As a result, we expect better free-carrier mobility in APNRs. This is consistent with monolayer phosphorene, in which the mobility along the armchair direction is impressive while it is smaller along the zigzag direction.[22, 23]

The dependence of the band gap versus the ribbon width through quantum confinement is a topic of particular interest when designing nanostructures.[15, 33-35] Figure 3 (a) indicates that the bands gaps for both types of PNRs exhibit significant dependences on their widths. However, their scaling laws are surprisingly different. The reduction of the band gap with increasing ribbon width occurs faster in APNRs than it does in ZPNRs such that an APNR will possess a smaller band gap than a ZPNR of equal width. This may provide a convenient tool for experiments to identify APNRs and ZPNRs with the similar geometric width. More interestingly, in Figure 3 (a), a power-law fit reveals that the scaling in APNRs perfectly follows the widely known $1/w^2$ relation while the scaling in ZPNRs follows a $1/w$ relation. The coexistence of vastly different scaling laws has not observed in other known 1D nanostructures, and merits special attention.

The dependence of PNR band gaps on their geometric widths can be explained in the context of quantum confinement and through considerations of well-studied 1D nanostructures. Previous studies of silicon nanowires reveal that their band gaps scale approximately as $1/w^2$.[33] This is attributed to the fact that the kinetic energy of a classical particle is proportional to the square of its momentum. On the other hand, GNRs are typified by Dirac Fermions, which obey a linear energy-momentum dispersion relation. Accordingly, GNRs exhibit the $1/w$ scaling law.[28, 34] These considerations indicate that electrons behave like classical particles along the confinement direction of APNRs, which is the zigzag direction, but behave like relativistic particles along the confinement direction of ZPNRs, which is the armchair direction.

The apparent division of classical-like and relativistic-like behaviors based on ribbon-direction is confirmed by the anisotropic band structure of monolayer phosphorene plotted in Figure 3 (c). Previous studies show that phosphorene has a highly anisotropic effective mass; it is heavy along the Γ-Y (zigzag) direction, but light along the Γ-X (armchair) direction.[22, 23] Upon further investigation, we find that these two directions differ by more than just band curvature; there are much deeper physical reasons. Away from their band edges, both the first valence and conduction bands are nearly linear across a wide energy range along the Γ-X direction, as seen from Figure 3 (c). Accordingly, in the right subfigure of Figure 3 (c), the valence band does not resemble a typical parabolic form, but rather is fitted remarkably well by the relativistic band dispersion as $E = \sqrt{m^2 c^4 + c^2 p^2}$, in which *c* is the Fermi velocity in phosphorene, *p* is the momentum and *m* is the rest effective mass. According to our fitting, the Fermi velocity of holes is around 8 x 10$^5$ m/s and the rest mass is small (*around 0.12 m$_0$*). These curved Dirac-Fermion dispersions are similar to those of bilayer graphene.[36, 37]

The observed Dirac-Fermion curves prompt a new understanding of the anisotropies of 2D phosphorene and their effects on its free carriers. The electrons and holes behave like classical particles along the armchair direction while they behave like relativistic particles along the zigzag direction. This explains the observed high mobility of the free carriers in few-layer black phosphorus;[17, 18] as in few-layer graphene, the relativistic Dirac-Fermion dispersion diminishes the electron-phonon coupling.[38, 39] The coexistence of classical and relativistic properties in few-layer black phosphorous may give rise to novel magnetotransport and optical properties, such as an unusual quantum Hall effect associated with coexistence of classical and relativistic particles, and anisotropic excitons.[21, 40]

The distinct scaling laws of differently-oriented phosphorene allow one to control the band gap and transport properties of PNRs. For example, APNRs exhibit a sensitive dependence of their band gaps on ribbon width, thus they can be used to realize a broad range of band gaps, more so than the less-sensitive ZPNRs. Devices that require uniform, consistently-performing components would incorporate ZPNRs, which would exhibit only small variations of the band gap due to edge/width fluctuations.

The effective masses of electrons and holes in PNRs, and their dependence on ribbon width, are presented in Figure 3 (b). The effective mass of holes in ZPNRs is fitted to the global band maximum instead of at the local maximum ($\Gamma$ point). Generally, ZPNRs has a much larger effective mass than APNRs. The effective masses of ZPNRs are generally independent on the ribbon width, while those of APNRs quickly decrease with an increasing ribbon width. In the limit of large ribbon widths, as shown in Figure 3 (c), their effective masses resemble their bulk limits and do not change much.

Figure 4 shows the wave functions at the top of valence bands at the $\Gamma$ points of both types of PNRs. Most wave functions are well confined within the ribbon, indicating that the edge state is eliminated by hydrogen passivation. Interestingly, the phase of the wave functions always alternates between neighboring P-P dimmers, which is similar to what is found in monolayer phosphorene. As a result, all of the atoms on a single edge have the same phase. The phases between two edges can be controlled by using the width of PNRs. For example, for N-ZPNRs, the two edges will have agreeing phases for even N and opposite phases for odd N. The 4-ZPNR structure shown in Figure 4 (a1) is significantly deformed by edge relaxations, skewing the wave function distribution. This is consistent with the small deviations observed in the band gap evolution shown in Figure 3 (a).

Moving beyond band structures, we present the DFT-calculated single-particle optical absorption spectra of PNRs. The optical absorption polarizability α is obtained by formulas in Refs. 41 and 42 to avoid the artificial impact of the vacuum between ribbons:

$$\alpha(\omega) = \frac{4e^2 A}{\omega^2} \sum_{k,c,v} |\vec{\lambda} \cdot \langle k,c|\vec{v}|k,v\rangle|^2 \delta[\omega - (\omega_{k,c} - \omega_{k,v})], \qquad (1)$$

which is obtained by multiplying the imaginary part of calculated dielectric susceptibility, $\kappa = (\varepsilon - 1)/4\pi$, by the cross-section area *A* of the supercell perpendicular to the nanoribbon orientation.

It must be noted that our calculated absolute absorption energy and detailed absorption profiles cannot be directly compared with experiments, because self-energy corrections and excitonic effects, which are known to be important for producing accurate optical spectra for low-dimensional semiconductors,[43] are not included. However, the general optical transition activities and symmetry-related selection rules of PNRs can still be understood within the single-particle picture.

The unusually flat valence band in ZPNRs, depicted in Figure 2 (a), gives rise to a large van Hove Singularity (vHS) and may be prominent for optical absorption. However, our simulations show that the flat band is not active for optical excitations in ZPNRs. This is consistent with the anisotropic optical absorption predicted in few-layer black phosphorus.[21] The optical transitions of monolayer phosphorene between the first pair of valence and conduction bands are not active for incident light polarized along the zigzag direction, which is exactly the orientation of ZPNRs. Meanwhile, we observe active higher-energy transitions in ZPNRs close to the edge of the first Brillouin zone, as marked by an arrow in Fig. 2 (a). This results from band folding. Therefore, ZPNRs have a dark symmetry gap that is similar to what has been observed in silicon nanowires.[42] We expect to find low photoluminescence (PL) in these ZPNRs, making them suitable for applications associated with long-lifetime dark excitons.[44] Additionally, this symmetry gap, combined with the larger comparative band gap, causes ZPNRs to have much higher absorption edge energies than do APNRs of similar widths.

For APNRs, optical transitions are active for the first pair of valence and conduction bands. The corresponding optical transitions are also marked in Figure 2 (b). As shown in Figure 5 (b), they exhibit the characteristic 1D vHS decay ($1/\sqrt{E}$), and the optical absorption edge starts from the band gap exactly. Band gap engineering can, therefore, be used to directly tune the optical absorption of APNRs. In particular, because of strongly active transitions around the band gap, these PNRs may exhibit excellent PL efficiency.

**CONCLUSIONS**

In this work, we employed first-principles simulations to study the electronic structures and optical responses of 1D semiconductor, PNRs. Our simulation shows that their band gaps can be substantially enhanced by quantum confinement. However, the scaling law of the band gap versus ribbon width is vastly different for different ribbon orientations. Further inspection reveals that this results from major anisotropic effects exhibited by phosphorene. The free carriers of phosphorene behave as relativistic particles along the armchair direction, giving rise to a $1/w$ scaling law in ZPNRs. On the other hand, the zigzag direction hosts classically-behaving particles, resulting in the $1/w^2$ scaling law for APNRs. Finally, we discussed the nature of the effective masses and optical responses of APNRS and ZPNRs. Although both ribbons exhibit a nearly direct band gap, ZPNRs have symmetry-gap forbidden transitions around the band gap while APNRs have strong optical excitations at the band edge.

*Acknowledgement:* This work is supported by the National Science Foundation Grant No. DMR-1207141. The computational resources have been provided by the Lonestar of Teragrid at the Texas

Advanced Computing Center (TACC). The first-principles calculation is performed with the Quantum Espresso package.[45]

*Conflict of Interest:* The authors declare no competing financial interest.

**Figures and captions:**

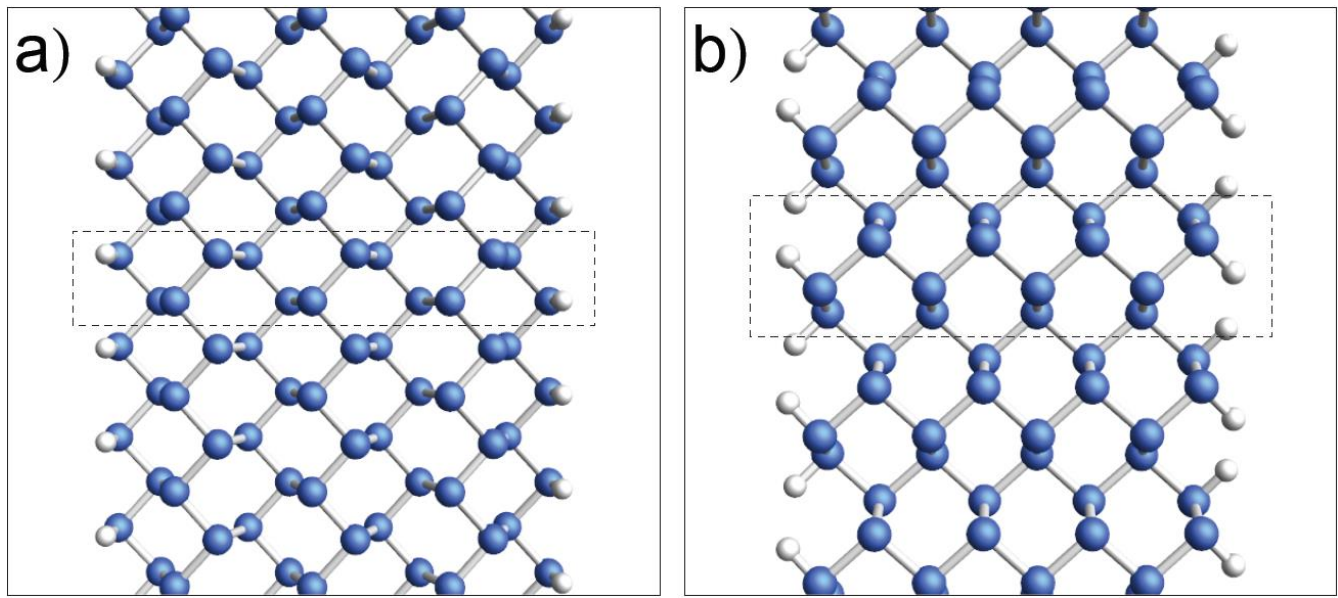

**Figure 1** (a) and (b) Top views of ball-stick models of atomic structures of 7-ZPNRs and 8-APNRs. The edge dangling bonds are passivated by hydrogen (white-colored) atoms. The unit cell is marked by dashed rectangles in (a) and (b).

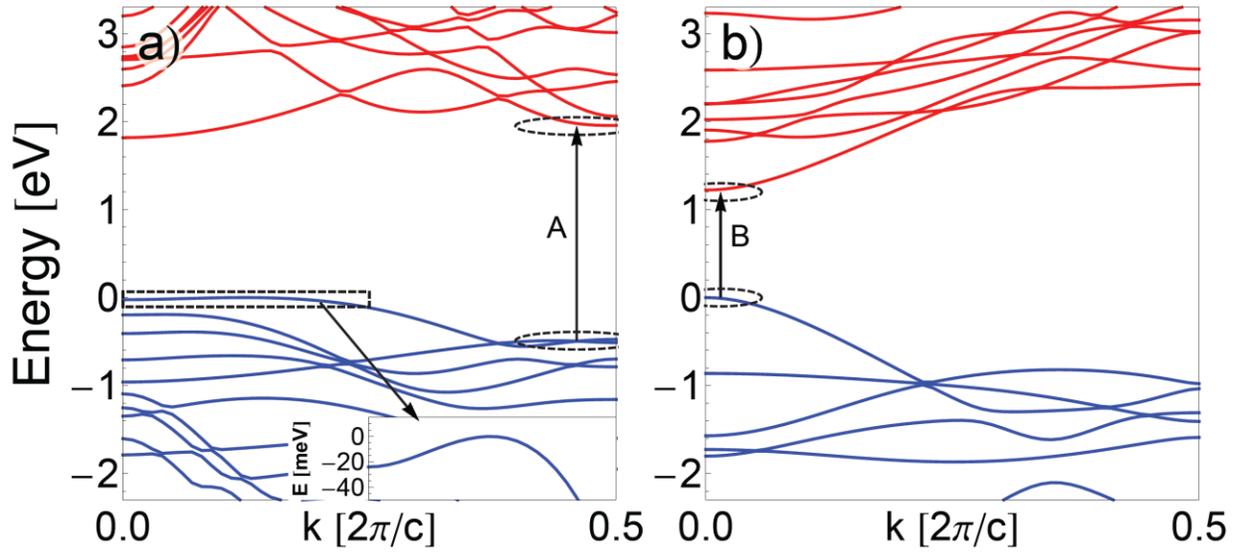

**Figure 2** (a) and (b) Band structures of 6-ZPNRs and 5-APNRs, respectively. The top of valence band is set to be zero in all plots. Valence bands are blue while conduction bands are red. The lowest-energy bright optical transition is marked as A and B in (a) and (b), respectively. The top of valence band of the 6-ZPNRs in (a) is highlighted by a dashed rectangle and amplified in the inset at the right-bottom corner.

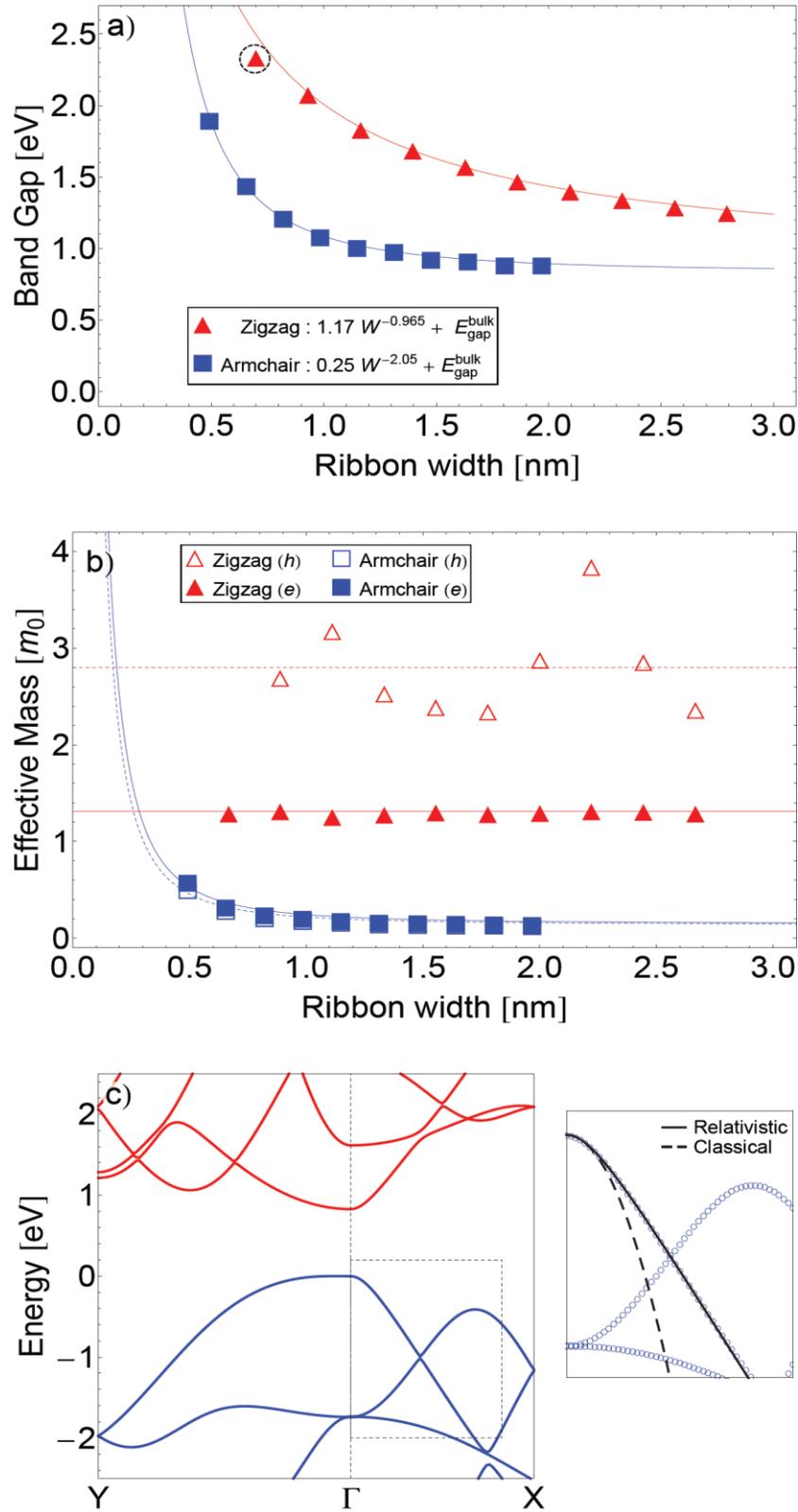

**Figure 3** (a) Band gap vs the width (*w*) of PNRs. The circle point is not included in the fitted curve because it is affected by significant deformation at the edges of an extremely narrow PNR. (b) Effective mass of electrons and holes vs the width (*w*) of PNRs. The fitted curve roughly follows a $1/w^2$ scaling law. (c) The band structure of monolayer phosphorene. The dashed rectangle is amplified in the right subfigure, in which the band energies are fitted with dispersion relations for a classical particle (dashed line) and a relativistic particle (solid line), respectively.

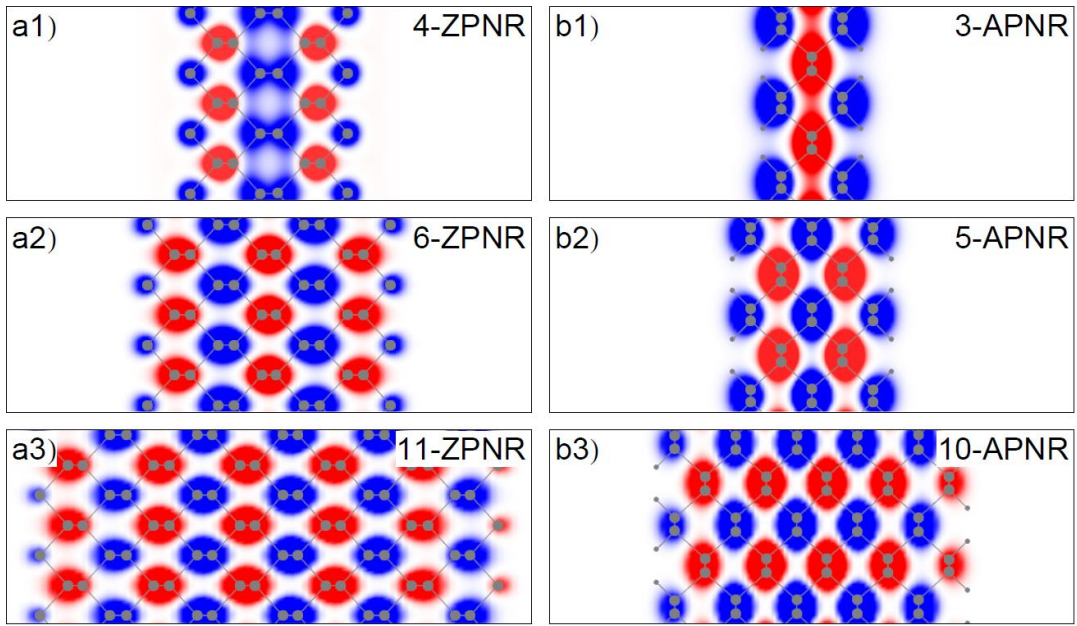

**Figure 4** (a1) to (a3) Top views of wave functions of the top of valence band at the $\Gamma$ point of ZPNRs. (b1) to (b3) Top views of wave functions of the top of valence band at the $\Gamma$ point of APNRs. The blue and red colors indicate the positive and negative phases of the wave functions.

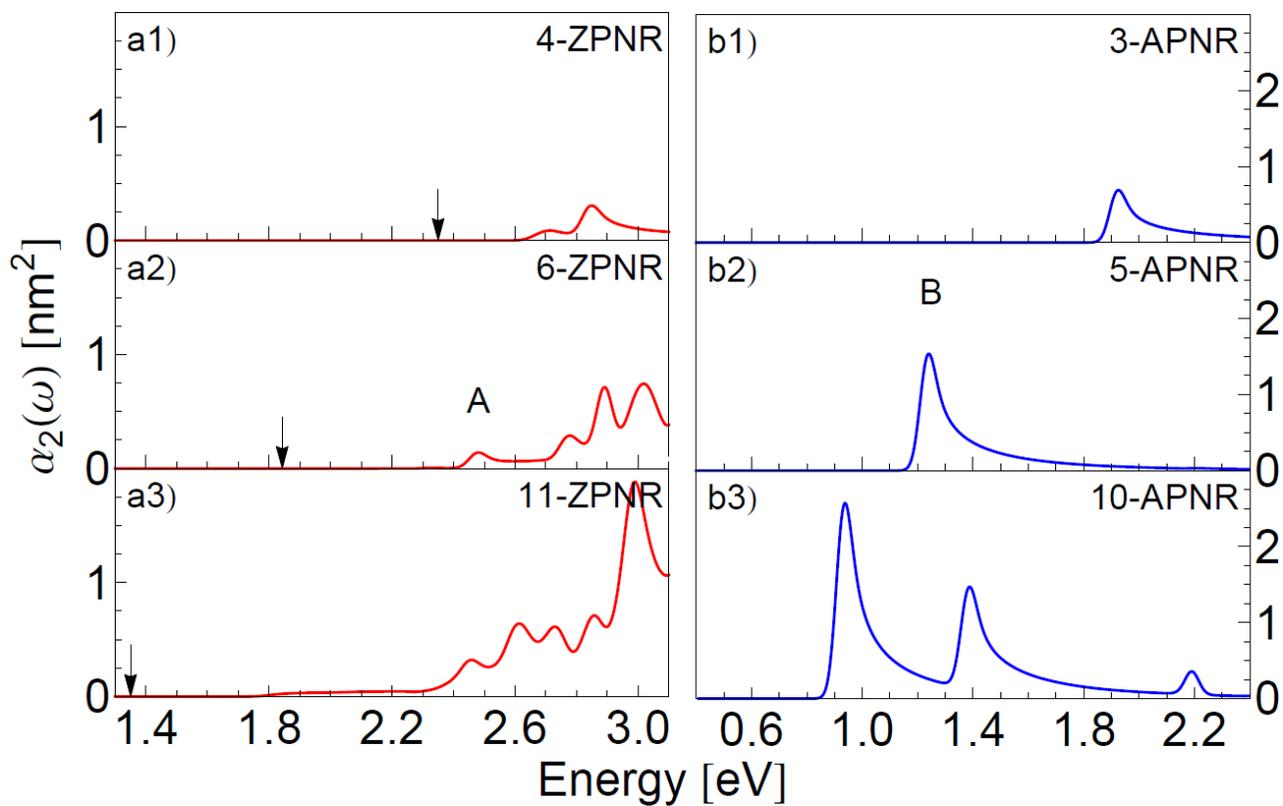

**Figure 5** (a1) to (a3) Single-particle optical absorption spectra of ZPNRs. (b1) to (b3) those of APNRs. For ZPNRs, the DFT-calculated band gap is marked by a black arrow. The transitions contributing to the peaks A and B in (a2) and (b2) are also marked in band-structure plots of Figures 2.